\documentclass[useAMS,usenatbib]{mn2e}

\title[Radio and $\gamma$-ray follow-up of PKS\,1510$-$089]{Radio and $\gamma$-ray follow-up of the exceptionally high activity state of PKS\,1510$-$089 in 2011} 
\author[M. Orienti, S. Koyama, F. D'Ammando et al. ]
  {M. Orienti$^{1,2}$\thanks{E-mail: orienti@ira.inaf.it},
S. Koyama$^{3,4}$, F. D'Ammando$^{2,5}$, M. Giroletti$^{2}$, M. Kino$^{3}$,
\newauthor H. Nagai$^{3}$, T. Venturi$^{2}$, D. Dallacasa$^{1,2}$,
G. Giovannini$^{1,2}$, E. Angelakis$^{6}$, 
\newauthor L. Fuhrmann$^{6}$, T. Hovatta$^{7}$, W. Max-Moerbeck$^{7}$, F. K. Schinzel$^{8}$,
K. Akiyama$^{3,4,9}$,  
\newauthor K. Hada$^{2}$, M. Honma$^{3,10}$, K. Niinuma$^{11}$, D. Gasparrini$^{12}$, T. P. Krichbaum$^{6}$,  
\newauthor I. Nestoras$^{6}$, A. C. S. Readhead$^{7}$, 
J. L. Richards$^{13}$, D. Riquelme$^{14}$, A. Sievers$^{14}$,
\newauthor H. Ungerechts$^{14}$, J. A. Zensus$^{6}$ \\
$^1$Dipartimento di Astronomia, Universit\`a di Bologna, via Ranzani 1,
I-40127, Bologna, Italy \\
$^2$INAF -- Istituto di Radioastronomia, via Gobetti 101, I-40129, Bologna,
Italy \\
$^{3}$National Astronomical Observatory of Japan 2-21-1 Osawa, Mitaka, Tokyo 181-8588, Japan\\
$^{4}$Department of Astronomy, Graduate School of Science, The University of Tokyo, 7-3-1 Hongo, Bunkyo-ku, Tokyo 113-0033, Japan\\
$^{5}$Dipartimento di Fisica, Universi\`a degli Studi di Perugia, I-06123 Perugia, Italy\\
$^{6}$Max-Planck-Institut f\"ur Radioastronomie, Auf dem H\"ugel 69,
53121 Bonn, Germany \\
$^{7}$Cahill Center for Astronomy and Astrophysics, California Institute
of Technology 1200 E. California Blvd., Pasadena, CA 91125, USA \\
$^{8}$Department of Physics and Astronomy, University of New Mexico, Albuquerque, NM 87131, USA\\ 
$^{9}$Research Fellow of the Japan Society for the Promotion of Science\\ 
$^{10}$Department of Astronomical Science, The Graduate University for
Advanced Studies, 2-21-1 Osawa, Mitaka, Tokyo 181-8588, Japan\\
$^{11}$Graduate School of Science and Engineering, Yamaguchi
University, 1677-1 Yoshida, Yamaguchi, Yamaguchi 753-8512, Japan\\
$^{12}$Agenzia Spaziale Italiana (ASI) Science Data Center, I-00044
Frascati (Roma), Italy\\
$^{13}$Department of Physics, Purdue University, 525 Northwestern Avenue, West
Lafayette, IN 47907, USA\\
$^{14}$Instituto de Radioastronom\'ia Millim\'etrica (IRAM), Avenida Divina Pastora 7, Local 20, 18012 Granada, Spain\\
}
\date{Received \today; accepted ?}

\pagerange{\pageref{firstpage}--\pageref{lastpage}} \pubyear{2002}

\def\LaTeX{L\kern-.36em\raise.3ex\hbox{a}\kern-.15em
    T\kern-.1667em\lower.7ex\hbox{E}\kern-.125emX}

\begin{document}

\label{firstpage}

\maketitle

\begin{abstract}
We investigate the radio and $\gamma$-ray variability of the flat
spectrum radio quasar PKS\,1510$-$089 in the time range 
between 2010 November and 2012 January.
In this period the source showed an intense activity, with two
major $\gamma$-ray flares detected in 2011 July and October.
During the latter episode both the $\gamma$-ray and the radio flux density
reached their historical peak. Multiwavelength analysis
shows a rotation of about 380$^{\circ}$ of the optical
  polarization angle close in time with the 
rapid and strong $\gamma$-ray
flare in 2011 July. An
enhancement of the optical emission and an increase of the fractional
polarization both in the optical and in radio bands is observed about three
weeks later, close in time with another $\gamma$-ray outburst. On the
other hand, after 2011 September a huge radio outburst has been
detected, first in
the millimeter regime followed with some time delay at centimeter down
to decimeter wavelengths. This radio flare is characterized by a
rising and a decaying stage, in agreement with the
formation of a shock and its evolution, as a consequence of expansion and 
radiative cooling. If the $\gamma$-ray flare observed in
2011 October is related to this radio outburst, then this strongly
indicates that the region
responsible for the $\gamma$-ray variability is not within the broad
line, but a few
parsecs downstream along the jet.

\end{abstract}


\section{Introduction}

Relativistic jets are one of the most powerful manifestations of the
release of energy produced by supermassive black holes in active
galactic nuclei (AGN). Their emission
is observed across the entire electromagnetic spectrum, 
from the radio band to $\gamma$ rays, and it is due to relativistic
particles producing synchrotron radiation in low energy bands (from
radio to optical/X-rays), which may also give rise to high-energy
emission by inverse Compton processes.
The discovery of $\gamma$-ray emission from 
relativistic jets in AGN represented a breakthrough in our understanding 
of the physical processes involved.  \\
Important clues on the jet physics have been obtained by the study of
the blazar population. Blazars are radio-loud AGN whose jet axis is closely
aligned with our line of sight. In these objects the jet emission is highly
amplified due to Doppler boosting effects, and a high level of
variability is detected at all wavelengths. In particular, blazars
with strong $\gamma$-ray emission are brighter and more luminous at
radio frequencies \citep{kovalev09}, have faster jets
\citep{lister09c}, and 
higher variability Doppler factors \citep{savolainen10} with respect to
those without significant $\gamma$-ray emission.\\ 
Despite decades of efforts, many aspects of the physics of
relativistic jets remain elusive. In
particular, the location and the mechanisms responsible for the high-energy
emission and the connection of the variability at different
  wavelengths
are among the greatest challenges in the study of AGN. Recently,
  observations performed with the Very Long Baseline Array (VLBA),
  which allow sub-milliarcsecond angular resolution,
  suggested that the region responsible for the high-energy emission
  is located downstream along the jet at a distance of a few parsecs 
from the central AGN \citep[e.g.][]{marscher08,larionov08}. 
On the contrary, following the causality argument, the
very short time scale variability detected at $\gamma$ rays
may imply that high energy emission is produced in a small region. If
the size of the emitting region is related to the jet cross sectional
radius, this suggests a location within the Broad Line Region (BLR, $<$ 1 pc) 
\citep[e.g.][]{tavecchio10}. 
If we consider that the entire cross section of the jet is responsible for the $\gamma$-ray emission,
this scenario is rather difficult to
reconcile with its location at a distance of several parsecs from the
central black hole (e.g. Sikora et al. 2008; Marscher et al. 2010), unless
the collimation angle of the jet is extremely small. 
For example, in the case of PKS\,1510$-$089 \citet{jorstad05} derived a half-opening angle of 0$^{\circ}$.2 and the corresponding radius of the jet cross section at about ten pc is $\sim$10$^{17}$ cm \citep{marscher10}.\\
Another explanation may involve turbulence in the jet flow.
\citet{marscher11} suggested that a
turbulent multi-zone model may be able to describe the
high-energy emission produced along the
jet, several parsecs away from the central engine. In this model, the
short $\gamma$-ray variability may be caused by a turbulent
jet plasma that passes through a standing shock along the jet. \\
Recent multiband studies of
blazars suggested that the high-energy emitting region is at parsec-scale
distances from the central black hole. For example, in the blazar OJ\,287
two major outbursts at millimeter wavelengths 
from a jet feature occurred almost
simultaneously with both optical and $\gamma$-ray flares, suggesting that
radio, optical and $\gamma$-ray variability is produced in a
single region along the jet located about 14 pc downstream from the
central engine \citep{agudo11a}. Further evidence for a
location of the high-energy variability at few parsecs from the broad
line region was provided by the study of
the multiwavelength outbursts in the blazars AO\,0235+164 and 3C\,345
\citep{agudo11b,schinzel12}. \\
Among the blazar population, the flat spectrum radio quasar (FSRQ) 
PKS\,1510$-$089 is one of the most
active objects where quiescent periods are interspersed with high
activity states with abrupt flux density increase at all wavebands
\citep[e.g.][]{venturi01,jorstad01,tornikoski94}. PKS\,1510$-$089 was
discovered as a $\gamma$-ray source by EGRET, and
it is one of the three FSRQs detected at very-high-energy (VHE, E $>$ 100
GeV) by MAGIC and H.E.S.S. (Cortina 2012, Wagner 2010). 
During the EGRET era the $\gamma$-ray emission of 
PKS\,1510$-$089 was found to be only slightly
variable \citep{hartman99}.
However, since 2008 the source has entered a very active period and
many rapid and intense flaring episodes have been detected by AGILE
and {\it Fermi}
\citep[e.g.][]{pucella08,dammando09,abdo10,dammando11}. 
Moreover, the high
variability level detected in $\gamma$ rays has been observed
across the entire electromagnetic spectrum.
Optical and radio outbursts seem to follow the high activity
  states in $\gamma$ rays with time delays of a few days in the
  optical, up to a few months as we consider the longer wavelengths of
the radio band \citep{abdo10}. Furthermore the strong $\gamma$-ray
flares observed in 2008 September and 2009 April are likely related to
the ejection of a new superluminal jet component
\citep{marscher10}. However, not all the $\gamma$-ray flares have
the same characteristics in the various energy bands
\citep[e.g.][]{dammando11} making the interpretation of the mechanism
responsible for the high-energy emission not trivial.\\
Of particular interest is the second half of 2011, when some intense
rapid $\gamma$-ray flares were detected. 
The first rapid flare was observed on
July 4 by AGILE and {\it Fermi} \citep{donnarumma11,dammando11b}.
The second episode started on October 17 and reached the maximum on
October 19 \citep{hungwe11}.
During the latter flare the source reached its highest $\gamma$-ray flux,
becoming the second brightest AGN ever observed by {\it Fermi}. \\
Triggered
by this extreme $\gamma$-ray activity, multifrequency radio VLBI and
single-dish observations
in the centimeter and millimeter bands were performed. In this paper we
investigate the connection of the $\gamma$-ray activity detected by the
{\it Fermi} Gamma-ray Space Telescope with the
emission at lower frequencies, focusing on the two main $\gamma$-ray
flares which occurred in 2011 July and October.\\
The paper
is organized as follows: in Section 2 we describe the radio
data from VERA, VLBA, Medicina, F-GAMMA, 
and OVRO observations.
In Section 3 we report the analysis of the {\it Fermi}-LAT
data. 
In Section 4 we present the results of the radio and $\gamma$-ray comparison, 
while discussion and concluding remarks
are presented in Sections 5 and 6, respectively.\\ 

Throughout this paper, we assume the following cosmology: $H_{0} =
71\; {\rm km \; s^{-1} \; Mpc^{-1}}$, $\Omega_{\rm M} = 0.27$ and
$\Omega_{\rm \Lambda} = 0.73$, in a flat Universe. 
At the redshift of the target, $z$ = 0.361 \citep{thompson90}, 
the luminosity distance $D_{\rm L}$ is
1913.2 Mpc, and 1 arcsec = 5.007
kpc. \\

\section{Radio data}

\subsection{VERA observations}

From 2010 November to 2012 January, PKS\,1510$-$089 was observed with
four VERA (VLBI Exploration of Radio Astrometry) 
stations at 22 GHz using the left hand circular polarization (LHCP)
feed only, typically twice per month, 
for a total of 32 observing epochs, within the framework of the
  Gamma-ray Emitting Notable-AGN monitoring by Japanese VLBI (GENJI)
  programme \citep[for more details see][]{nagai12b}. 
Observations of PKS\,1510$-$089 were
spread into several scans of about 5 minutes each, for a total
on-source observing time of about 15--30 minutes for each run. \\
Data reduction was performed using the NRAO's Astronomical Image
Processing System (\texttt{AIPS}). {\it A-priori} amplitude calibration was
derived with the \texttt{AIPS} task APCAL 
on the basis of the measurements of the system temperatures 
and the antenna gain information for each VERA antenna. Uncertainties
on the amplitude calibration are within 10\%. The source is
strong enough to allow the fringe fitting with a solution interval of
one minute to preserve the phase coherence. Final images were produced
after a number of phase self-calibration iterations. The flux density
was derived by means of the \texttt{AIPS} task JMFIT which performs a Gaussian
fit on the image plane. The typical resolution is about
(1.5$\times$1.0) milliarcsecond. Total intensity flux densities are reported in 
Table \ref{vera}.\\
 
\subsection{MOJAVE data}

We investigated the pc-scale morphology and 
flux density variability at 15 GHz by means of 14-epoch VLBA
data from the MOJAVE programme \citep{lister09}. The datasets span the
time interval between 2010 November and 2012 March. In addition, 
we included also
observations performed in 2012 April and May, 
in order to better characterize the
proper motion of the jet components likely ejected close in time with
the $\gamma$-ray flares.
We imported the calibrated {\it uv}-data into the NRAO \texttt{AIPS}
package. 
For a proper comparison with the VERA data, 
in addition to the {\it full-resolution} images, we produced
also {\it low-resolution} images in total intensity 
considering the same {\it uv}-range of
the VERA data (i.e. from 50 to 170 M$\lambda$) and the same resolution.
To derive the polarization information we also produced
Stokes' Q and U images. The flux density of
the core region
was derived by means of the \texttt{AIPS} task JMFIT which performs a Gaussian
fit on the image plane. Total intensity flux density and polarization
information are reported in Table \ref{mojave}.\\

\subsection{Medicina observations}

Since 2011 July, after the $\gamma$-ray flare detected by {\it
  Fermi}-LAT \citep{dammando11b}, PKS\,1510$-$089 has been monitored
almost once per month 
with the Medicina single-dish telescope at 5 and 8.4 GHz.
Observations have been performed with the new Enhanced Single-dish
Control System (ESCS), which provides enhanced sensitivity and supports
observations with the cross scan technique. At each frequency the
typical on source time is 40 seconds and the flux density was
calibrated with respect to 3C\,286. The flux densities at 5 and 8.4 GHz
measured with the Medicina telescope are listed in Table \ref{medicina}. \\

\subsection{F-GAMMA observations}

The cm/mm radio light curves of PKS\,1510$-$089 have been obtained within the 
framework of a {\sl Fermi} related monitoring program of
$\gamma$-ray blazars (F-GAMMA program, Fuhrmann et al. 2007, Angelakis
et al. 2008). The millimeter observations are closely coordinated with
the more general flux monitoring conducted by IRAM, and data from both 
programs are included in this paper. The
overall frequency range spans from 2.64\,GHz to 142\,GHz using the 
Effelsberg 100-m and IRAM 30-m telescopes and observations are
performed roughly once per month.\\
The Effelsberg measurements were conducted with the secondary focus
heterodyne receivers at 2.64, 4.85, 8.35, 10.45, 14.60, 23.05, and 32.0\,GHz.  
The observations were performed quasi-simultaneously with cross-scans, 
that is slewing over the source position, in azimuth and elevation
direction with adaptive number of sub-scans for reaching the desired
sensitivity \citep[for details see][]{fuhrman08,angelakis08}.
Pointing off-set correction, gain correction, 
atmospheric opacity correction and
sensitivity correction have been applied to the data.\\

The IRAM 30-m observations were carried out with calibrated
cross-scans using the new EMIR horizontal and vertical polarization receivers
operating at 86.2 and 142.3\,GHz. The opacity corrected intensities 
were converted into the standard temperature scale and finally
corrected for small remaining pointing offsets and systematic
gain-elevation effects. The conversion to the standard flux density scale
was done using the instantaneous conversion factors derived from 
frequently observed primary (Mars, Uranus) and secondary 
(W3(OH), K3-50A, NGC\,7027) calibrators.

\subsection{OVRO observations}

PKS\,1510$-$089 is part of an ongoing blazar monitoring program at 15 GHz, the Owens Valley
Radio Observatory (OVRO) 40-m radio telescope. This
monitoring program includes over 1500 confirmed and candidate $\gamma$-ray loud
blazars above declination $-20^{\circ}$ \citep{richards11}. The sources in this
program are observed in total intensity twice per week with a 4~mJy
(minimum) and 3\% (typical) uncertainty. Observations are performed
with a dual-beam (each 2.5~arcmin FWHM) Dicke-switched system using
cold sky in the off-source beam as the reference. Additionally, the
source is switched between beams to reduce atmospheric variations. The
absolute flux density scale is calibrated using observations of
3C\,286, adopting the flux density (3.44~Jy) from \citet{baars77}. 
This results in about a 5\% absolute scale uncertainty, which is not
reflected in the plotted errors. \\

\section{{\it Fermi}-LAT Data: Selection and Analysis}
\label{FermiData}

\begin{figure}
\begin{center}
\includegraphics{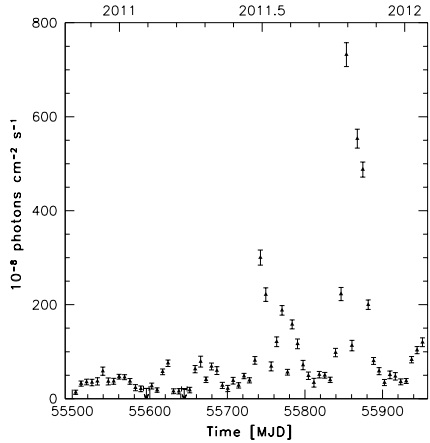}
\vspace{6.5cm}
\caption{{\it Fermi}-LAT integrated flux (E $>$ 100 MeV) light curve of
PKS\,1510$-$089 obtained from 2010 November 1 (MJD 55501) to 2012 January
31 (MJD 55957) using 1-week time bins. Arrows refer to 2-$\sigma$
  upper limit on the source flux. Upper limits are computed when TS $<$ 10.}
\label{LAT1}
\end{center}
\end{figure}

\begin{figure}
\centering
\includegraphics{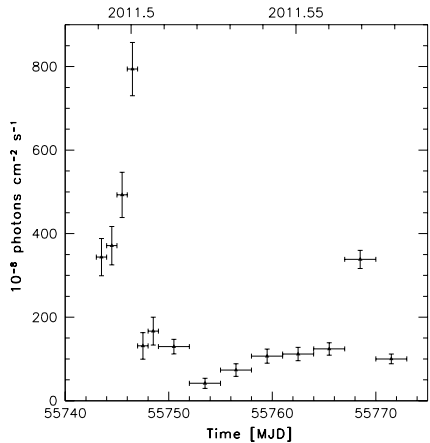}
\vspace{6.5cm}
\caption{Integrated flux (E $>$ 100 MeV) light curve of PKS\,1510$-$089
  obtained from 2011 July 1 to 30 (MJD 55743--55773) with either 1-day
  (close to peak) or 
  3-day time bins.}
\label{LAT2}
\end{figure}

\begin{figure}
\begin{center}
\includegraphics{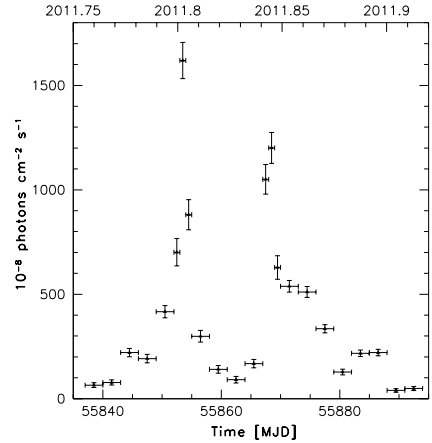}
\vspace{6.5cm}
\caption{Integrated flux (E $>$ 100 MeV) light curve of PKS\,1510$-$089
  obtained from 2011 October 3 to 2011 November 28 (MJD 55837--55893) 
with either 1-day
  (close to the peak) or 3-day time bins.}
\label{LAT3}
\end{center}
\end{figure}

The Large Area Telescope (LAT) on board {\it Fermi} 
is a $\gamma$-ray telescope operating from $20$\,MeV to
$>300$\,GeV. The instrument is an array of $4 \times 4$ identical towers, each
one consisting of a tracker (where the photons are pair-converted) and a
calorimeter (where the energies of the pair-converted photons are
measured). The entire instrument is covered with an anticoincidence detector
to reject the charged-particle background. The LAT has a large peak effective
area ($\sim$ $8000$\,cm$^2$ for $1$\,GeV photons), an energy resolution
typically about $10\%$, and a field of view (FoV) of about $ 2.4$ \,sr
with an angular resolution ($68\%$ containment angle) better than 1$^{\circ}$ for
energies above $1$\,GeV. Further details about the LAT are given by
\citet{atwood09}.\\ 
The LAT data reported in this paper were collected over 15 months of {\it
  Fermi} operation, from 2010 November 1 (MJD 55501) to 2012 January 31 (MJD
55957). During this time the LAT instrument operated almost entirely in survey
mode. The analysis was performed with the \texttt{ScienceTools} software
package version v9r23p1. The LAT data were extracted within a $15^{\circ}$
Region of Interest (RoI) centered at the radio location of
PKS\,1510$-$089. 
Only events belonging to the `Source' class were used. In
addition, a cut on the zenith angle\footnote{The zenith angle is
  defined as the angle of a photon's apparent origin to the
  Earth-spacecraft vector.} ($< 100^{\circ}$) was also applied to
reduce contamination from the Earth limb $\gamma$ rays, which are produced by cosmic rays interacting with the upper atmosphere. The spectral analysis (from which we
derived spectral fits and photon fluxes) were performed with the post-launch instrument response functions (IRFs) \texttt{P7SOURCE\_V6} using an
unbinned maximum likelihood method implemented in the Science tool
\texttt{gtlike}.\\ 
The background model used to extract the $\gamma$-ray signal includes a Galactic diffuse emission component and an isotropic component. The model that
we adopted for the Galactic component is given by the file gal\_2yearp7v6\_v0.fits, and the isotropic component, which is the sum of the extragalactic diffuse
emission and the residual charged particle background, is parametrized by the file
iso\_p7v6source.txt\footnote{http://fermi.gsfc.nasa.gov/ssc/data/access/lat/Background\\Models.html}. The
normalizations of both components in the background model were allowed
to vary freely during the spectral point fitting. \\ 
We examine the significance of the $\gamma$-ray signal from the sources by means of the Test Statistics (TS) based on the likelihood ratio test. The Test
Statistic TS = 2$\Delta$log(likelihood) between models  with and without the source is a measure of the probability to having a $\gamma$-ray source at the
localization specified, which compares models whose parameters have
been adjusted to maximize the likelihood of the data given the model \citep{mattox96}. The source model used in \texttt{gtlike} includes all the point sources from the second {\it Fermi}-LAT catalogue (2FGL; Nolan et al. 2012) that fall within 20$^{\circ}$ from PKS\,1510$-$089.
In addition in the model we included also
the FSRQ TXS\,1530$-$131, at $6^{\circ}$ from the source, detected in flare by
{\it Fermi}-LAT on 2011 August 22 \citep{gasparrini11}. The spectra of these
sources were parametrized by power-law functions, except for
2FGL\,J1504.3$+$1029, for which we used a log-parabola for its
spectral 
modeling as
in the 2FGL catalogue. We removed from the model the sources having test
statistic TS $<$ 25 and/or fluxes below 1.0$\times$10$^{-8}$ photons cm$^{-2}$s$^{-1}$ over 15 months and repeated the fit. Thus a final fitting procedure has been performed with the sources within 10$^{\circ}$ from PKS\,1510$-$089 included with the normalization factors and the photon indices left as free parameters. For the sources located between 10$^{\circ}$ and 15$^{\circ}$ we kept the normalization and the photon index fixed to the values obtained in the previous fitting procedure.
The RoI model includes also sources at distances between 15$^{\circ}$ and
20$^{\circ}$ from the 
target source, which can contribute to the total counts observed in the RoI due to the energy depended size of the point spread function of the instrument. For these
additional sources, normalizations and indices were fixed to the values of
the 2FGL catalogue. \\
Following the 2FGL catalogue the spectral model used for PKS\,1510$-$089
is a log-parabola, 
dN/dE~$\propto$~E/E$_{0}^{-\alpha-\beta\,\rm log(E/E_0)}$ \citep[]{landau86, massaro04}, where the parameter $\alpha$ is the spectral
  slope at the energy E$_0$ and the parameter $\beta$ measures the curvature around the peak. We fixed the reference energy E$_0$ to 259.6
  MeV as in the 2FGL catalogue. The fit over the entire period 2010
  November--2012 January (MJD 55501--55957)
  in the 0.1--100 GeV energy range results in a TS = 20678, with
  $\alpha = 2.22 \pm$0.02, $\beta = 0.07 \pm$ 0.01, and an integrated average
flux of (88.3 $\pm$ 1.4) $\times$10$^{-8}$ photons cm$^{-2}$
s$^{-1}$. Using a power-law model the fit yielded to a TS = 20155,
with a photon index $\Gamma = 2.33 \pm$0.02 and an integrated average flux of (90.1 $\pm$ 1.3) $\times$10$^{-8}$
photons cm$^{-2}$ s$^{-1}$, corresponding to an isotropic $\gamma$-ray
luminosity of $\sim$1.6$\times$10$^{47}$ erg s$^{-1}$. As a comparison
the isotropic
$\gamma$-ray luminosity over the first two years of {\it Fermi} operation is
$\sim$1.8$\times$10$^{47}$ erg s$^{-1}$, indicating that the source activity
remained quite high throughout the {\it Fermi} era.\\ 
Fig. \ref{LAT1} shows the {\it Fermi}-LAT $\gamma$-ray light curve
of PKS\,1510$-$089 during the 15 months considered in this paper using 1-week time
bins and the log-parabola spectral model. For each time bin the $\alpha$ and
$\beta$ parameters were frozen to the value resulting from the likelihood
analysis over the entire period. If TS $<$ 10 the value of the fluxes were replaced by the 2-$\sigma$ upper limits. The systematic uncertainty in
the flux is energy dependent: it amounts to $10\%$ at 100 MeV, decreasing to
$5\%$ at 560 MeV, and increasing to $10\%$ above 10 GeV \citep{ackermann12}.\\
Several prominent $\gamma$-ray peaks are clearly visible in the 1-week light
curve over 15 months. We produced two additional light curves focused on
the periods when the largest flares occurred: 
2011 July 1--30 (MJD 55743--55773) and
2011 October
3--November 28 (MJD 55837--55893), using the log-parabola spectral
model with either 1-day (close to the peaks, corresponding to the time
bins with higher statistics) or 3-day time bins
(Fig.~\ref{LAT2} and Fig.~\ref{LAT3}). The 2011 July and October flares are
characterized by a doubling time scale ($t_{\rm var}$) of 2 days and 1
day, respectively.
The daily peak isotropic $\gamma$-ray luminosities are
1.6$\times$10$^{48}$ erg s$^{-1}$ on July 4, 3.7$\times$10$^{48}$ erg s$^{-1}$
on October 19, and 2.3$\times$10$^{48}$ erg s$^{-1}$ on November 2, with an
increase of a factor of 10, 23, and 14, with respect to the average value. To
calculate these values we used the photon index obtained from a power-law
model estimated over the relative 1-week time bin: $\Gamma_{\rm July}
= 2.21 \pm$0.05, $\Gamma_{\rm October} = 2.09 \pm$0.03, and
$\Gamma_{\rm November} = 2.26 \pm$0.04, respectively.

\begin{figure}
\begin{center}
\includegraphics{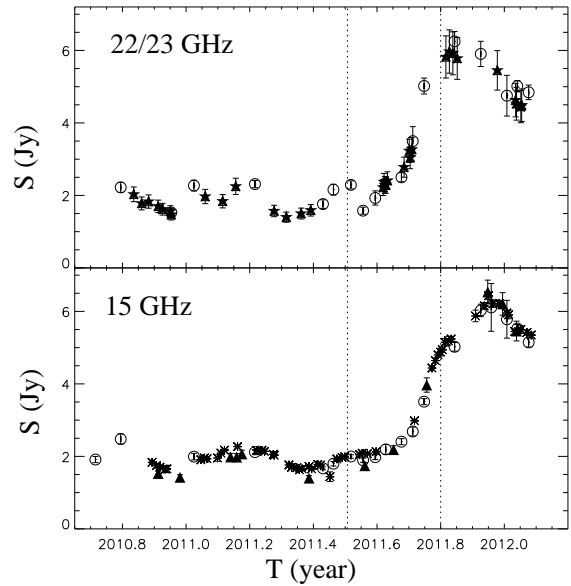}
\vspace{10cm}
\caption{Radio light curves at 22 and 23 ({\it upper panel}) and 15 GHz ({\it
    lower panel}) of PKS\,1510$-$089. Empty circles are F-GAMMA data at 15 and
  23 GHz; filled stars are VERA data at 22 GHz, asterisks and filled triangles
  represent OVRO and MOJAVE data at 15 GHz. Vertical
  lines indicate the time of the $\gamma$-ray flares.}
\label{radio_curva}
\end{center}
\end{figure}

\begin{figure}
\begin{center}
\includegraphics{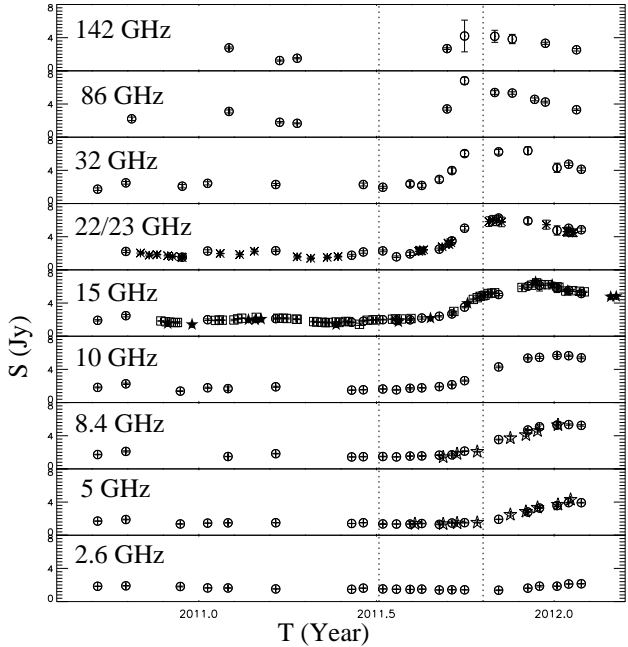}
\vspace{10cm}
\caption{Radio light curves of PKS\,1510$-$089 between 1 November 2010
  and 31 January 2012. Circles, asterisks, squares, filled stars, and
  empty stars refer to F-GAMMA data, 22-GHz VERA data, 15-GHz OVRO
  data, 15-GHz MOJAVE data, and Medicina data respectively. Vertical
  lines indicate the time of the $\gamma$-ray flares.}
\label{multi_ovro}
\end{center}
\end{figure}

\begin{table}
\caption{22-GHz flux density of the core component 
measured by the VERA interferometer.}
\begin{center}
\begin{tabular}{lc|lc}
\hline
Obs.\ date& $S_{\rm 22 GHz}$&Obs.\ date& $S_{\rm 22 GHz}$\\
          &  Jy           &          & Jy\\
\hline
&&&\\
2010/11/01& 2.03& 2011/08/16& 2.29\\ 
2010/11/10& 1.78& 2011/08/17& 2.29\\ 
2010/11/18& 1.83& 2011/08/19& 2.41\\ 
2010/11/29& 1.70& 2011/09/07& 2.78\\ 
2010/12/04& 1.62& 2011/09/13& 3.23\\ 
2010/12/13& 1.56& 2011/09/14& 3.04\\ 
2010/12/14& 1.47& 2011/09/16& 3.27\\ 
2011/01/22& 1.97& 2011/10/25& 5.82\\ 
2011/02/11& 1.84& 2011/10/29& 5.97\\ 
2011/02/26& 2.25& 2011/11/02& 5.92\\ 
2011/04/11& 1.57& 2011/11/07& 5.78\\ 
2011/04/25& 1.40& 2011/12/23& 5.45\\ 
2011/05/12& 1.50& 2012/01/13& 4.63 \\
2011/05/23& 1.59& 2012/01/14& 4.53\\ 
2011/08/14& 2.22& 2012/01/19& 4.45 \\
2011/08/15& 2.37& 2012/01/20& 4.49 \\ 
&&&\\
\hline
\end{tabular}
\label{vera}
\end{center}
\end{table}

\begin{table}
\caption{Flux density and polarization information of the central
  region of PKS\,1510$-$089 from the MOJAVE 15-GHz VLBA data.}
\begin{center}
\begin{tabular}{lccc}
\hline
Obs.\ date& $S_{\rm 15 GHz}$& $S_{\rm p}$& $\chi$\\
          &  mJy           & mJy ( \% )& deg\\
\hline
&&&\\
2010/11/29& 1527& 31  (2.0\%)& 75 \\
2010/12/24& 1422& 16  (1.1\%)& 65 \\
2011/02/20& 1984& 13  (0.7\%)& 50 \\
2011/02/27& 1980& 14  (0.7\%)& 50 \\
2011/03/05& 2067& 23  (1.1\%)& 85 \\
2011/05/21& 1399& 10  (0.7\%)& 45 \\
2011/07/24& 1741& 29  (1.7\%)& 45 \\
2011/08/26& 2186& 26  (1.2\%)& 40 \\
2011/10/03& 3969& 20  (0.5\%)& 55 \\
2011/12/12& 6536& 93  (1.4\%)& 65 \\
2011/12/29& 6207&172  (2.8\%)& 60 \\ 
2012/01/14& 5458&167  (3.0\%)& 55 \\
2012/03/04& 4802& 98  (2.0\%)& 50\\
2012/03/27& 4758&128  (2.7\%)& 24\\
&&&\\
\hline
\end{tabular}
\label{mojave}
\end{center}
\end{table}

\begin{table}
\caption{Results of the Medicina 32-m radio telescope.}
\begin{center}
\begin{tabular}{lcc}
\hline
Obs.\ date& $S_{\rm 5 GHz}$& $S_{\rm 8.4 GHz}$\\
\hline
&&\\
2011/07/12&   -          & 1.30$\pm$0.05\\
2011/08/10& 1.38$\pm$0.05&      -       \\
2011/09/08& 1.36$\pm$0.11& 1.44$\pm$0.11\\
2011/09/22& 1.46$\pm$0.07& 1.82$\pm$0.20\\
2011/10/13& 1.53$\pm$0.05& 2.08$\pm$0.05\\
2011/11/16& 2.49$\pm$0.17& 3.72$\pm$0.22\\
2011/12/01& 2.85$\pm$0.15& 4.10$\pm$0.15\\
2011/12/13& 3.34$\pm$0.12& 4.59$\pm$0.15\\
2012/01/03& 3.70$\pm$0.20& 5.30$\pm$0.40\\
2012/01/17& 4.30$\pm$0.20&  - \\
&&\\
\hline
\end{tabular}
\label{medicina}
\end{center}
\end{table}

\section{Results}

\subsection{Light curves}
\label{sec_flux}

In 2011, PKS\,1510$-$089 showed high activity in the $\gamma$-ray energy
band, with
an average weekly flux between 100 MeV and 100 GeV of about 
8$\times$10$^{-7}$~photons~cm$^{-2}$~s$^{-1}$
(Fig. \ref{LAT1}). The second half of 2011 is
characterized by two major flares which occurred 
 in July (Fig. \ref{LAT2}) and
October (Fig. \ref{LAT3}).\\  
To compare the flux variability at high
($\gamma$ rays) and low (radio band) energies, we analyzed
millimeter/centimeter radio observations spanning a time interval between 
2010 November and 2012 January, i.e. long enough to constrain any flux
density variation well before and after the main $\gamma$-ray
outbursts. In Fig. \ref{radio_curva} we plot the radio light curves at
23 and 15 GHz, i.e. those frequencies with the best time sampling. The light
curves are characterized by subsequent increase and decrease of the
flux density. 
Since 2011 September the flux density at both 15 and 23 GHz abruptly
increased, reaching its maximum at the beginning of November for the
23 GHz, followed with one-month delay at 15 GHz, when the peak was
observed around middle of December. 
It is worth noting that the flux density on parsec scales
derived by VLBI data, i.e. VERA data at 22 GHz and VLBA data
at 15 GHz (triangles and stars in Fig. \ref{radio_curva}) strictly
follow the single-dish flux density trend, indicating that the
flux density variability is dominated by the parsec-scale emitting region.\\
In Fig. \ref{multi_ovro} we report the multifrequency light curves from
2.6 to 142 GHz, in order to compare the flux density behaviour at various
frequencies.
The abrupt flux density increase reported since the
beginning of September \citep{nestoras11} is firstly detected above 22
GHz. At these frequencies the light curves show a structure similar to
a plateau starting from the time of the October $\gamma$-ray flare.
At lower frequencies the flux density increase is much smoother and
with some time delay \citep{mo11b}.\\

\subsection{Time delay}
\label{sec_delay}

\begin{figure}
\begin{center}
\includegraphics{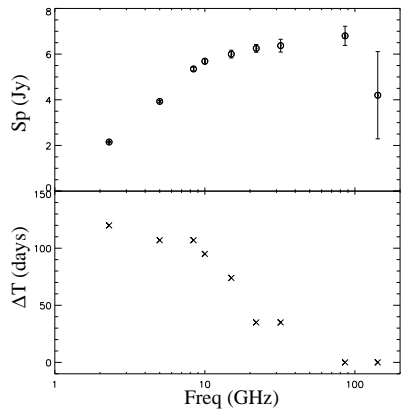}
\vspace{6cm}
\caption{Peak flux density ({\it top}) and the time
  delay between burst maxima ({\it bottom}), normalized at the
    86-GHz values, against the observing frequency.} 
\label{peak_delay}
\end{center}
\end{figure}

\begin{figure}
\begin{center}
\includegraphics{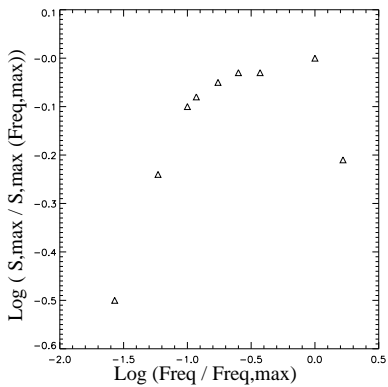}
\vspace{6cm}
\caption{Observed maximum relative flux density scaled to the
  frequency at which the maximum occurs.}
\label{peak_norm}
\end{center}
\end{figure}

Statistical studies of the light curves of blazars showed that 
the maximum flux density value reached during an outburst strictly 
depends on the observing frequency and usually it is not
simultaneous at the various frequencies \citep[e.g.][]{hovatta08,hughes11}.  
In the shock scenario we expect that the peak occurs first in the
millimeter regime then followed with some time delay at the longer
wavelengths, as a consequence of jet opacity.
Time lags between
frequencies are thus important to constrain the properties of the shock at
the origin of the observed outburst.\\
The main problem in determining the characteristics of the light curves 
depends on the time sampling of the observations at each
frequency: if observations are too sparse in time, the epochs
at which the flux density starts to rise and then peaks
cannot be constrained with adequate accuracy, 
causing large uncertainties on the
parameters of the shock. This is particularly critical at
high frequencies where the variability is faster and the atmosphere
is more incoherent, making the measurements more difficult and with
larger uncertainties.\\ 
The analysis of the multifrequency light curves clearly indicates that
at the highest frequencies, 86 and 142 GHz, the flux density 
reaches its maximum almost simultaneously at the end of September, 
while at longer wavelengths it occurs after some time delay 
(Fig. \ref{multi_ovro}). In
Fig. \ref{peak_delay} we report the peak flux density ({\it upper panel})
and the time lag ({\it lower panel}), normalized at the 86-GHz
  values, as a function of the frequency.
The time delay increases with the wavelengths, from about one month at 
32 and 22 GHz, up to
several months in the decimeter regime (at 2.6 GHz). Although the
peak at 86 and 142 GHz seems to precede the high $\gamma$-ray
flare detected in October, the actual peak time could have been
missed during observations due to the sparse time coverage. For this
reason, we cannot exclude a simultaneity between the millimeter peak
and the $\gamma$-ray flare. 
It is worth noting that the flux density
increase observed since 2011 September might be related to the
$\gamma$-ray flare detected in 2011 July. However, flares at millimeter
wavelengths often peak quasi-simultaneously with the $\gamma$-ray
flare \citep{abdo10}, and
a delay of almost three
months at millimeter wavelengths is unusual. As a comparison, we
  note that during the high activity states showed by PKS\,1510$-$089
  in the first half of 2009, the peak at 230 GHz occurred about one
  month after the $\gamma$-ray flare observed in 2009 March, and almost
  simultaneously with the $\gamma$-ray flare detected in 2009 April \citep{abdo10}.\\
Following the approach discussed by \citet{valtaoja92} 
we computed the maximum relative flux
density scaled to the frequency at which the maximum occurs
(Fig. \ref{peak_norm}). Since the maximum peak flux density seems to
be at 86
GHz, we normalized the values to this frequency. In the shock scenario
we expect three different evolutionary epochs: 
1) the growth stage at high frequencies, where the radiation is
optically-thin, 2) a plateau stage, where the turnover frequency moves
towards lower frequencies while the peak flux density is almost
constant, 3) a decay stage at low frequencies, where the peak flux
density is lower than in the plateau stage due to strong energy
losses. To study the evolutionary stages of the outburst that took
place in 2011 October, we fit the three parts of the 
normalized spectrum with a power-law (Fig. \ref{peak_norm}). We
find a slope of 0.5, and -0.9 for the declining (between 10 and 5 GHz)
and the
rising part (between 142 and 86 GHz) respectively, and 0.06 for the
plateau (between 86 and 15 GHz). 
The slope of the declining part has been
computed considering the 5 GHz as the lowest frequency, since at 2.6
GHz the flux density is still increasing in 2012 January. 
The rising part is not
well-constrained due to the poor time sampling and 
the large uncertainty on the flux density. \\

\subsection{Parsec-scale properties}
\label{sec_parsec}

\begin{figure}
\begin{center}
\includegraphics{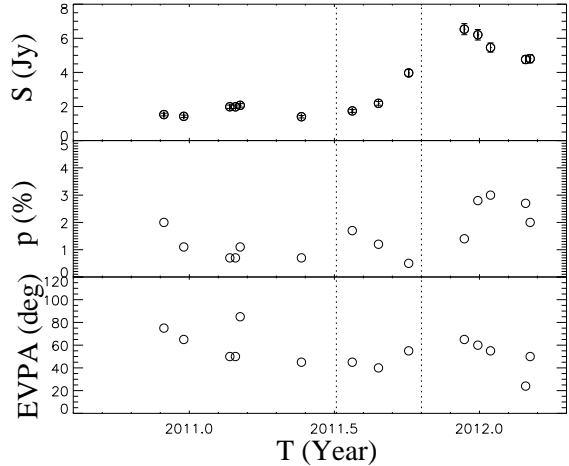}
\vspace{8cm}
\caption{Total intensity and polarization properties
  of the core region of PKS\,1510$-$089 at 15 GHz. From top to bottom panel:
  total intensity flux density; polarization
  percentage; polarization angle. Vertical lines
  indicate the time of the $\gamma$-ray flares.}
\label{multi_polla}
\end{center}
\end{figure}

\begin{figure}
\begin{center}
\includegraphics{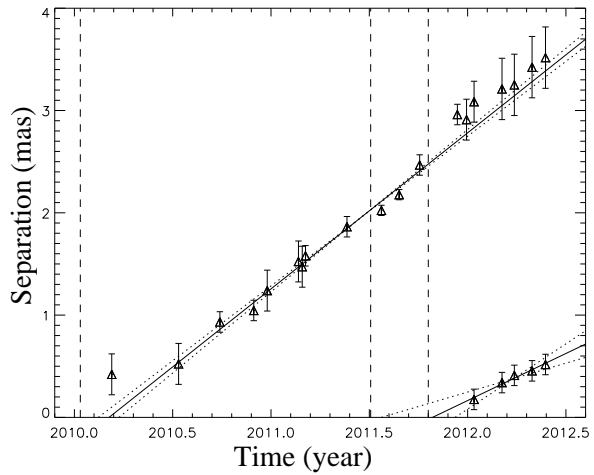}
\vspace{6cm}
\caption{Changes in separation with time between components C,
  considered stationary, 
  J, and the new knot ejected in 2011. 
  The solid line represents the regression fit to the 15-GHz VLBA
  data, while the dashed lines represent the uncertainties from the
  fit parameters. Dashed vertical lines indicate the time of the $\gamma$-ray
flares.} 
\label{moto}
\end{center}
\end{figure}

Single-dish observations cannot
separate the various contributions to the emission 
and the flux density variability
originating in the central region, as
well as its polarization properties, may be washed out. 
Observations with parsec-scale resolution are required to disentangle the
contribution of the core region from the emission arising from the jet
and extended features. \\
On the
parsec scale, the radio emission from PKS\,1510$-$089 is dominated by the
core component, from which the jet emerges with a position angle of
$\sim-35^{\circ}$, 
i.e. to the north-west direction (Fig. \ref{fig_mojave}). 
Fig. \ref{multi_polla} shows the radio properties of the parsec-scale core
region obtained by means of high-resolution VLBA observations at
15 GHz. As pointed out in Section \ref{sec_flux} the variability
shown by the source clearly originates in the core region. The core
light curve shows a strong increase starting from
2011 September. Simultaneously, the polarization percentage of the core
decreases, while the polarization angle remains almost constant. After
the $\gamma$-ray flare in 2011 October, the fractional polarization 
increases from $\sim$0.5\% of October, with a polarized
flux density of 20 mJy, to $\sim$3.0\% measured
in 2012 January, when the polarized emission reached $\sim$170 mJy. On
the other hand, no significant changes were observed in 
the electric vector position angle (EVPA), which ranges between
55$^{\circ}$ and 65$^{\circ}$ in the same period. \\
The analysis of the multi-epoch parsec-scale morphology of
PKS\,1510$-$089 points out that the jet component, labelled J in
Fig. \ref{fig_mojave}, is moving outward from the core. To determine the
proper motion of the jet component we model-fitted
the visibility data  
of each observing epoch with Gaussian components
using the model-fitting option in \texttt{DIFMAP} 
(for a description of the model-fitting
procedure and the associated errors see Orienti et al. 2011b). 
We performed the study of the
source structure using 15-GHz VLBA data from the MOJAVE programme, 
since both the $uv$-coverage and
the sensitivity of the 22-GHz VERA data are not adequate.\\  
The angular separation velocity
and time of zero-separation ($T_{0}$) from the core were derived by means of a
linear fit. In addition to the 16 MOJAVE epochs considered in this
paper, we also fit earlier epochs already published in
\citet{orienti11}. From this analysis we found that component J,
labelled N3 in \citet{orienti11}, is separating from the core with an
angular velocity of 1.5$\pm$0.1 mas/yr, which corresponds to an
apparent velocity of (33.4$\pm$2.2)$c$ (Fig. \ref{moto}). 
This value is larger than what was found by \citet{orienti11} based on
five epochs only. The availability of additional 16 observing epochs allow
a more accurate estimate of the proper motion. 
The time of zero-separation
is 2010.18$\pm$0.05, in agreement with the previous
estimate. Interestingly, the ejection of this component is close
  in time with a $\gamma$-ray flare detected by AGILE on 2010 January \citep{striani10}.\\
The modelfit of the MOJAVE data sets did not reveal the ejection of a
new component until 2012 January. After this time the observations show 
the presence of   
a new feature that is moving away from the core. A linear fit of the
five epochs 
in which the new component is detected (i.e. from 2012 January
to May), provides an angular separation rate of 0.92$\pm$0.35
mas/yr, which corresponds to an apparent separation velocity of
(20.5$\pm$7.8)$c$. From the regression fit we estimate that the time of
zero-separation occurred about 2011.83 (i.e. October 26), making the
ejection of the blob close in time with the $\gamma$-ray flare
detected in 2011 October. However, the large
uncertainties due to the availability of a few epochs spanning a
  short time range do not
allow us to accurately constrain the precise time of zero separation,
which ranges between 2011.56 (i.e. July 23) and 2011.93 (i.e. December
5). Although the typical apparent speed measured for PKS\,1510$-$089
are usually larger than 15$c$ supporting the idea that the new jet
component is related to the October $\gamma$-ray flare, we cannot
exclude that its ejection occurred close in time with the 2011 July
outburst.\\

\subsection{Optical emission}

\begin{figure}
\begin{center}
\includegraphics{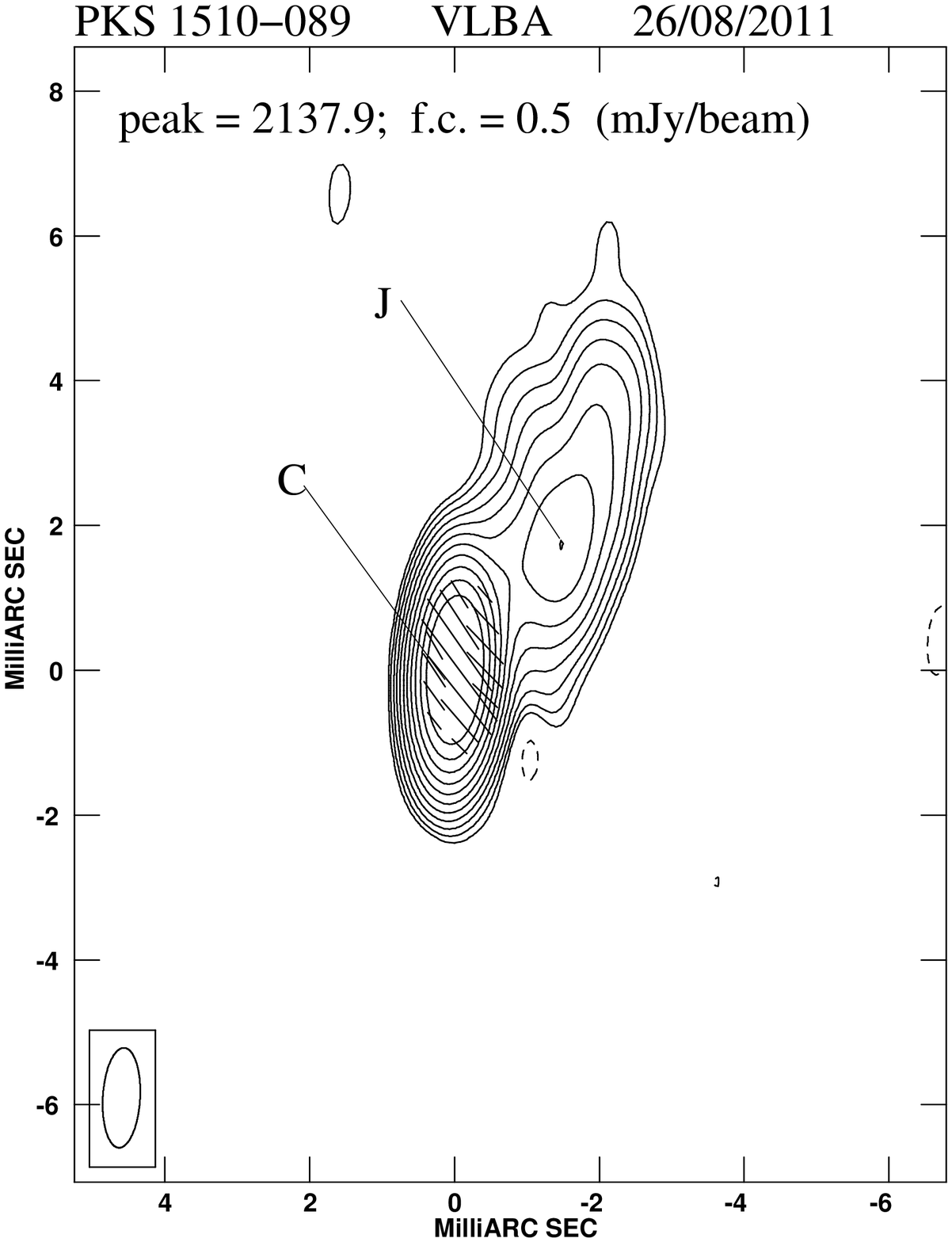}
\includegraphics{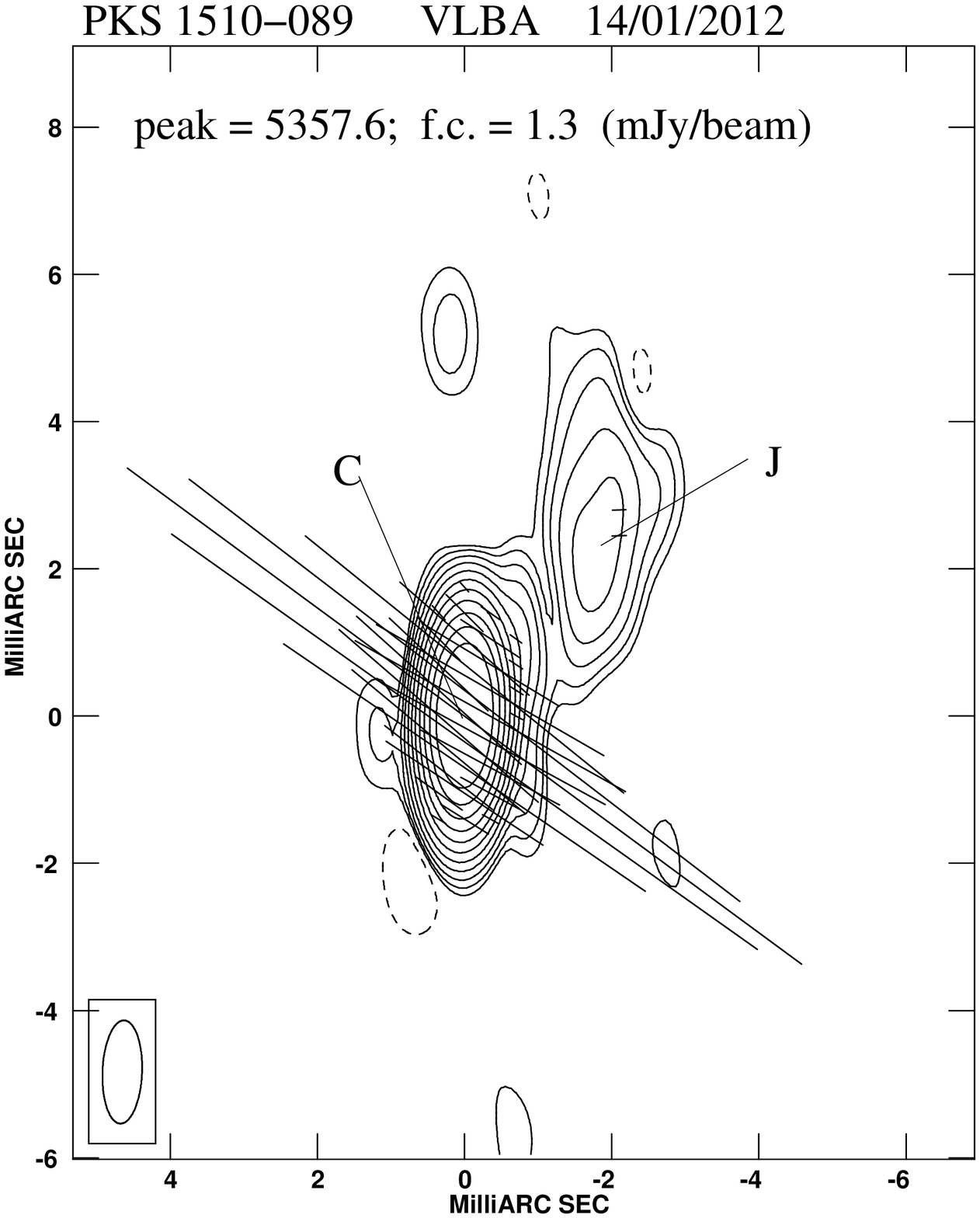}
\vspace{15cm}
\caption{15-GHz VLBA images of PKS\,1510$-$089 relative to 2011 August
  ({\it upper panel}) and 2012 January ({\it bottom panel}).
On each image, we provide the 
telescope array and the observing date; 
the peak flux density is in mJy/beam and the first contour (f.c.) 
intensity is in mJy/beam, which corresponds to three times 
the off-source noise level. Contour levels increase by a factor of 2. 
The restoring beam is plotted in the bottom left-hand corner. 
For the data sets with polarization information, the vectors 
superimposed on the I contours show the position angle of the E
vector, where 1 mm length corresponds to 14.3 mJy/beam. }
\label{fig_mojave}
\end{center}
\end{figure}

\begin{figure}
\begin{center}
\includegraphics{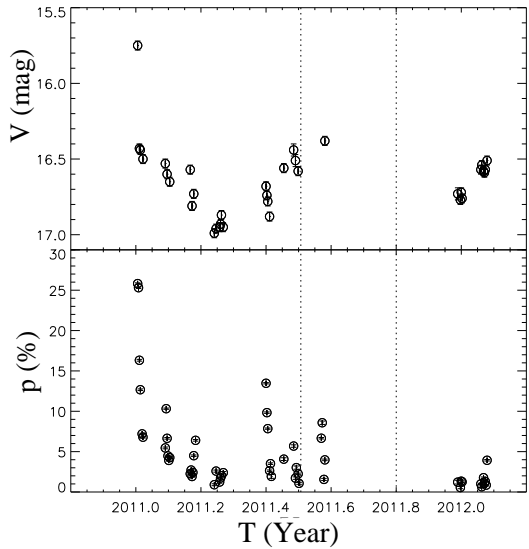}
\vspace{8cm}
\caption{Optical V-band light curve ({\it top}) and polarization
  percentage ({\it bottom}) of PKS\,1510$-$089 between 2011 January and
  2012 January. Vertical lines indicate the time of the
    $\gamma$-ray flares.}
\label{multi_ottico}
\end{center}
\end{figure}

\begin{figure}
\begin{center}
\includegraphics{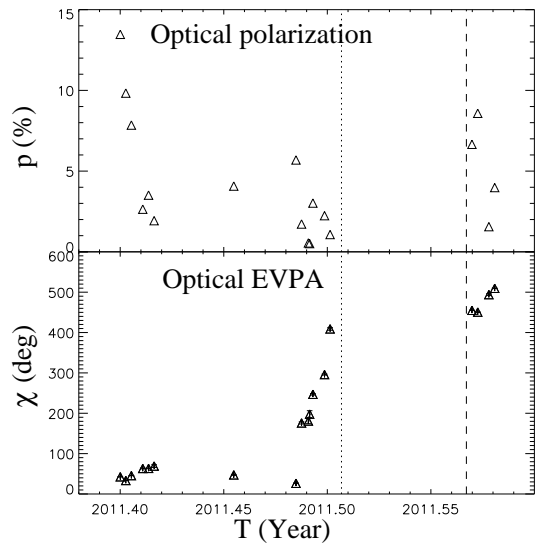}
\vspace{8cm}
\caption{Optical polarization percentage ({\it top}) and
  electric-vector position angle of the optical polarization
  ($\lambda$ = 500 -- 700 nm) of PKS\,1510$-$089 between 2011 May and
  2011 July. Vertical lines indicate the time of the
    strong $\gamma$-ray flare of 2011 July 4 ({\it dotted line}) and
    the R-band flare ({\it dashed line})
    detected between 2011 July 24 -- 26 \citep{hauser11}. Multiples of
180$^{\circ}$ are added to EVPA as needed to minimize the jumps in
consecutive values. } 
\label{angolo_ottico}
\end{center}
\end{figure}

Previous multifrequency studies of PKS\,1510$-$089 
showed an irregular flux variability between $\gamma$-ray and
optical emission. \citet{dammando11} found that during the high
activity of 2009 March, the optical outburst occurred with about 1
day lag with respect to the $\gamma$-ray flare. On the other hand, 
a lag of 13$\pm$1 days between the $\gamma$-ray and the optical peaks
was reported in \citet{abdo10} for the 2009 January and April
flares. Furthermore, in the same period the strong $\gamma$-ray flare
that took place in 2009 April seems to coincide with the end of a 50 day
rotation of the optical polarization angle, suggesting a connection
between these two energy bands \citep{marscher10}.\\
To test possible connection between the $\gamma$-ray flaring
events with changes in the optical emission we make use of the optical
data from the Steward Observatory blazar monitoring program of the
University of
Arizona\footnote{http://james.as.arizona.edu/$\sim$psmith/Fermi}. A
description of this monitoring project, the calibration and the data
products can be found in \citet{smith09}.\\
In Fig. \ref{multi_ottico} we show the V-band light curve and the
polarization percentage ($\lambda = 500-700$ nm). The mean V-band
magnitude is 16.65, while the mean polarization percentage is
$\sim$5\%. No obvious trend between the optical emission and
the fractional polarization has been found.\\
Interestingly, no significant increase in the optical luminosity has been
detected just after the $\gamma$-ray
flare observed in 2011 July. 
At the beginning of July the magnitude was $\sim$16.6 
and the polarized percentage was $\sim$1\%, well below the mean
value. This result is in
agreement with observations in R-band performed a few days after the
$\gamma$-ray flare, where the optical emission (R$\sim$16.1) 
is consistent with a
low activity state \citep{bachev11}. However, the $\gamma$-ray flare
occurred close in time with the end of a 7-day period during which the
optical polarization vector  
rotates of about 380$^{\circ}$ (Fig. \ref{angolo_ottico}). The
beginning of the rotation of the optical polarization vector coincides
with a local maximum in the percentage of polarization, which was
about 5.7\%, while the V-band magnitude was 16.4. The polarization
percentage dropped during the rotation, and it slightly increased at the end
of the period. The lack of observations after July 2 
do not allow us to describe the polarization trends
soon afterward the $\gamma$-ray flare.\\
An increase in both the
optical emission and fractional polarization is detected a few weeks
later, at the end of July, when the V magnitude reached 16.4. The polarized
emission increased up to 8.6\% and the optical polarization angle
rotates of about 60$^{\circ}$ in a 4-day period.
This optical outburst was more prominent in the R-band, where
the source magnitude reached 15.3 \citep{hauser11}. 
This event occurred close in time with an enhancement of the $\gamma$-ray
flux (Fig. \ref{LAT2}), and an increase of the radio polarization
percentage, while the radio polarization angle does not show
significant variation remaining stable between 40$^{\circ}$ and 60$^{\circ}$ 
(Fig. \ref{fig_mojave}). \\ 
Unfortunately, during the huge $\gamma$-ray flare occurred in 2011 October
the source was not observable in the optical and no information is
available.\\
It is interesting to note the presence of an optical flare at the
beginning of 2011 January, when the V-band emission reached a
magnitude of 15.75 and the fractional polarization was about
25\%. Then both the optical emission and polarization decreased
reaching a V-band magnitude of about 16.5 and a polarization percentage of
6.8\%. During this 7-day period the optical polarization angle
rotates of about 40$^{\circ}$. Although
no high $\gamma$-ray activity was detected in this period, the optical
flare is close in time with a local maximum in the 22-GHz light curve
(Fig. \ref{radio_curva}). \\

\section{Discussion}

Multiwavelength monitoring
campaigns suggest a relation between $\gamma$-ray flares and 
the radio variability,
explained in terms of a shock moving along the jet, whose
manifestation is a superluminal knot observable with
high-frequency VLBI observations
\citep[e.g.][]{marscher85,valtaoja92,hughes11}. The shock
model implies: 1) a growth stage, when the shock forms up to the development
of its maximum, which is observed not simultaneously at the various
energy bands due to opacity effects (Compton losses dominate); 2) a plateau, when energy losses and
gains are balanced (synchrotron losses dominate); 3) 
a decaying stage, when the shock fades due
to energy losses (adiabatic losses dominate). 
However, not all the outbursts, even produced in the same
source, behave similarly. For example, not all the $\gamma$-ray flares
show a counterpart in the various energy bands.
A clear example is
represented by 3C\,279 whose prominent $\gamma$-ray flare detected in
2009 February, and associated
with changes in the optical flux and polarization angle, 
is not related to any variability in the radio
band, even considering some time delay \citep{abdo10b}. A similar case
is represented by the misaligned object 3C\,84 where two major
$\gamma$-ray flares detected by {\it Fermi}-LAT and MAGIC do not have
an apparent counterpart in the radio band \citep{nagai12}.\\
The FSRQ PKS\,1510$-$089 is one of those objects which does not show a
trivial connection between the light
curves behaviour at the various frequencies during different outbursts. \\

\subsection{The high activity state in 2011}

In the second half of 2011, PKS\,1510$-$089 became very active, with
two main flaring episodes occurring in July and October, 
with a daily isotropic peak luminosity above 10$^{48}$
erg~s$^{-1}$. The $\gamma$-ray light curve clearly shows the abrupt
increase of the flux associated with the two prominent flares, both
characterized by short time variability $t_{\rm var}$. Contrary
to what was observed in the case of the FSRQ 3C\,454.3
\citep[e.g.][]{abdo11c}, the $\gamma$-ray flares of PKS\,1510$-$089 
are not preceded by 
a plateau in the $\gamma$-ray light curve, 
confirming the trend observed during previous
flaring episodes for this source. \\
On the basis of the short time
variability observed, we constrain the intrinsic size of the
$\gamma$-ray emitting region $R$ by means of the causality argument:\\

\begin{equation}
R \leq c t_{\rm var} \frac{\delta}{(1+z)}     
\label{size}
\end{equation}

\noindent where $\delta$ is the relativistic Doppler factor and $z$ is
the redshift. Assuming $\delta=20$, as typically found for this source
\citep[e.g.][]{dammando09,abdo10}, and $t_{\rm var}$ of 2 days and 1 day
for the flares observed in 2011 July and 2011 October, respectively,
(see Section 3), we obtain a size of
$R \leq$ 7.6$\times$10$^{16}$ cm and
$R \leq$ 3.8$\times$10$^{16}$ cm for the former and latter flare,
respectively. These values are in agreement with what was found by the
analysis of the spectral energy distribution during the high activity
states between 2008 and 2009 \citep{dammando09,abdo10}. \\
The analysis of the data from the multifrequency monitoring
campaigns obtained in the period considered in this paper 
showed that high activity was also
observed at centimeter and millimeter wavelengths. In particular, the
good time sampling at 22/23 GHz allowed us to identify several high
activity states, one almost simultaneous with the 2011 July
$\gamma$-ray flare, when the flux density increased by about 60\%
with respect to the minimum value measured in 2011
April. At lower radio frequencies no significant flux density increase
has been detected, while the lack of millimeter observations between
2011 June and 2011 September does not
allow us to investigate the variability at higher frequencies. 
Interestingly, the 2011 July $\gamma$-ray flare seems to occur at the
end of a 7-day period in which the optical EVPA rotates of
about 380$^{\circ}$. A similar behaviour was shown by 3C\,279 during
the strong $\gamma$-ray flare detected in 2009 February, which was
accompanied by an abrupt change of the optical EVPA, while in radio no
significant variability was observed \citep{abdo10b}.  \\
An optical outburst was observed
almost three
weeks after the 2011 July $\gamma$-ray flare, when another 
high activity $\gamma$-ray state was
observed (Fig. \ref{LAT2}). This outburst is also accompanied by an
increase of both optical and radio polarization percentage, and a
rotation of the optical EVPA of about 60$^{\circ}$ in 4 days. On the
contrary, no significant change in the radio EVPA has been noticed. \\
The most interesting feature shown by PKS\,1510$-$089 is the huge 
radio flare observed
since the beginning of 2011 September. Its multifrequency study indicates a
rising stage, first detected above 23 GHz, 
and a decaying stage, as expected in the general shock model
\citep[e.g.][]{valtaoja92}. The exceptional $\gamma$-ray flare detected in
  2011 October and the maximum in the millimeter light curves occur
  close in time, suggesting that the radio and $\gamma$-ray emission
  originates in the same region.\\
If the onset of the millimeter
outburst is a consequence of the formation of a
shock, then the $\gamma$-ray flare detected in October
took place when the shock
already moved downstream along the jet. The distance $\Delta r$ between
the region where the shock formed and the site responsible for the
$\gamma$-ray emission may be determined by means of:

\begin{equation}
\Delta r = \frac{\beta_{\rm app} c}{{\rm sin} \theta} \frac{\Delta t_{\rm
    obs}}{1+z} ,
\label{eq_distanza}
\end{equation}

\noindent where $\beta_{\rm app}$ is the apparent jet velocity, $c$ is
the speed of light, 
$\Delta t_{\rm obs}$ is the time elapsed between
the onset of the millimeter outburst and the $\gamma$-ray flare in the
observer's frame, and $\theta$ is the viewing angle \citep{pushkarev10}. 
In this case $\Delta t_{\rm obs}$ is about 40
days: $\Delta t_{\rm obs}$ is computed from the 22 GHz data between
2011 September 9, i.e. when the flux density doubled its
value, and 2011 October 17, i.e. the detection of the $\gamma$-ray flare.
If we assume
$\beta_{\rm app}$ = 25.4 (see Section 4.3) 
we find that the $\gamma$-ray flare is produced at a projected
distance of $\sim$ 0.6 pc (i.e. 0.1 mas at the source redshift) from
the site where the shock detected in radio band 
was formed. If we consider $\theta$ = 3$^{\circ}$, as derived from
previous studies of this source
\citep[e.g][]{marscher10,orienti11,lister09b,dammando11,abdo10}, 
we find that the de-projected
distance is about 10 pc. This result is in agreement with the idea
that at least some $\gamma$-ray flares do not take place 
within the broad line region.
The parsec-scale distance of the site
responsible for the $\gamma$-ray variability may be reconciled with
the small size of the emitting region derived from Eq. \ref{size}, in
the case the high-energy emission is due to turbulent plasma crossing
a pre-existent standing shock, likely the radio core, located along the jet at several parsecs
away from the nucleus \citep[e.g][]{marscher11}. 
In this case,
a contribution of Synchrotron Self-Compton (SSC) in addition to
inverse Compton produced by scattering of the infrared photons from the dusty
torus \citep[e.g.][]{sikora08} may be at
the origin of the $\gamma$-ray emission
\citep[e.g.][]{marscher11,marscher10,lahteenmaki03}. \\
The multi-wavelength variability of PKS\,1510$-$089 during
2011 has some similarities with the radio-to-$\gamma$-ray outburst of
the blazar BL Lacertae observed in 2005 \citep{marscher08}. Although
the time sampling of both the radio and optical observations are not
adequate for an accurate interpretation of the phenomena, we may speculate a
possible connection between the two main $\gamma$-ray flares
and the radio outburst of PKS\,1510$-$089 in
2011, as it was done for BL Lacertae. \\
In both sources the first high-energy flare coincides with an abrupt
rotation of the optical EVPA and the lack of significant flux density
variability at the longer radio wavelengths. The rotation of the
polarization angle suggests that the shock is likely produced by a
disturbance of the flow at the beginning of the jet. The perturbed flow
follows a spiral path as it propagates through the toroidal magnetic
field of the initial part of the jet, i.e. the acceleration and
collimation zone \citep{komissarov07}. 
A support to this interpretation comes from the
decrease of the polarization percentage during the rotation, when the
mean magnetic field of the disturbance is roughly transverse to the
field of the unperturbed flow. The lack of significant radio
counterpart indicates that this flare is taking place close to the
central region of the AGN, where the radio emission is
self-absorbed. As the perturbed flow propagates downstream the jet the
opacity decreases, and the outburst becomes visible in the radio
regime starting from the millimeter band down to 
longer wavelengths. The second high-energy flare would take place when
the knot encounters a standing
conical shock during its propagation along the jet. The perturbed flow
would be compressed by its passage through
the shock front. This would result in an amplification of the magnetic
field and an enhancement of its emission. 
As the knot continues to
propagate it becomes visible as a superluminal jet component
detectable with the high-resolution VLBA observations. 
In this scenario both gamma-ray flares would be produced by the same population of particles in two distinct moments (i.e. at distinct distances from the 
central engine) by means of different mechanisms.\\

\subsection{The long-term variability of PKS\,1510$-$089}

The flaring activity shown by PKS\,1510$-$089 since the launch of AGILE
and {\it
  Fermi}-LAT has been characterized
by several outbursts with different properties.
In the period
2008 January--April, when
intense activity was detected in the
$\gamma$-ray, optical and millimeter regimes, 
no similar trend was
found at the centimeter wavelengths \citep{dammando09}. 
A similar behaviour was found by \citet{dammando11} during the high
activity period at the beginning of 2009 March. 
Furthermore,
during the flaring episodes occurred in 2009 January, the optical
emission remained in a weak state \citep{marscher10}.\\
The flare
detected in 2011 July, characterized by a higher $\gamma$-ray luminosity
with respect to those previous
outbursts, occurred close in time with a significant rotation of the
optical EVPA, similar to what was observed in 2009 April. At 23 GHz there
is a hint of flux density increase, but the lack of observations at
higher frequencies does not allow us to reliably correlate the
$\gamma$-ray variability with the light curves at lower frequencies,
suggesting that opacity effects are dominant. \\
A different situation emerges from 
the strong $\gamma$-ray
flares in 2008 September and 2009 April, which seem strictly related to 
the ejection of superluminal jet knots and increase of the flux
density at high radio frequencies \citep{marscher10,orienti11}. 
The strong flare in 2009 April 
is also associated with an exceptional optical flare, when it reached
its historical peak with an R-band magnitude of about 13.6
\citep{larionov09}, and a 
large rotation of the optical and radio polarization
angle, indicating a common origin for the variability observed across
the entire electromagnetic spectrum. 
This flaring episode is similar to the radio-$\gamma$-ray behaviour 
shown by the flare in 2011 October, when the initial phase of
  the radio outburst seems to
precede the high energy variability. 
However, these episodes do not share all the same
characteristics. Although the lack of optical information does not
allow a complete multiband comparison, the polarimetric properties
derived suggest a different behaviour between these two flares. In the
first case the increase of the radio flux density is accompanied by a
drop in the fractional polarization and a rotation of about
90$^{\circ}$ of the EVPA. This behaviour is in agreement with a shock
propagating perpendicular to the jet axis \citep{orienti11}. 
On the other hand, in 2011 October the
polarization percentage at 15 GHz reaches a minimum just before the $\gamma$-ray
flare, and then increases as the emission switches from
optically-thick to optically-thin at
this frequency. However, the EVPA changes by only 20$^{\circ}$,
  which is difficult to reconcile with the propagation of a transverse
shock, but can be reproduced by an oblique shock.
In this case, the expected variations in the 
polarization angle are strongly related to 
the obliqueness of the shock itself and to the characteristics of the
magnetic field of the flow like its order and strength
\citep{hughes11}. \\

\section{Conclusions}

In this paper we presented results of the radio-to-$\gamma$-ray
monitoring of PKS\,1510$-$089 from 2010 November to 2012 January. 
Since 2011 July the source
became very active at high energies, reaching its
historical peak in $\gamma$ rays in 2011 October.\\ 
A multifrequency analysis showed
that the rapid and strong $\gamma$-ray flare detected in 2011 July 
is related to a rotation of the optical polarization angle
suggesting a common region responsible for both $\gamma$-ray and
optical emission. The lack of a simultaneous
increase of the flux density in the centimeter regime suggests that
the emitting region is close to the central AGN, at the beginning of
the jet, where the radiation is opaque at the radio wavelengths.\\
On the other hand, the strong flare in 2011 October seems to be
related to the huge radio outburst detected since the beginning of
2011 September 
starting from the higher radio frequencies. In this case the radio
outburst seems to precede the $\gamma$-ray flare, suggesting that the
site responsible for the $\gamma$-ray emission is located along the
jet, about 10
parsec away from the central engine. 
This strong flare seems also
related to the ejection of a new superluminal jet component from the
radio core. \\
As in the case of BL Lacertae, both $\gamma$-ray flares may be
interpreted by means of a single disturbance originated in the very
initial part of the jet, opaque to the radio wavelength. As the
perturbed flow propagates downstream the jet, the opacity decreases
and the variability becomes visible at longer wavelengths. As the flow
passes through a standing conical shock, that may be the radio core,
its emission is amplified and a second high-energy flare is
produced. In this case the dominant emission mechanism may be either inverse
Compton of infrared photons of the dusty torus or synchrotron
photons from the standing shock. However, a synchrotron self-Compton
origin cannot be ruled out. Then the
disturbance continues its way downstream the jet becoming visible as a
superluminal jet component by means of observations with
the milliarcsecond resolution reached by the Very Long Baseline
Interferometry (VLBI) technique.\\
It is worth noting that during 2011 the $\gamma$-ray and optical light
curves present additional high-activity states that complicate
this simple picture. Furthermore, the long term monitoring of
PKS\,1510-089 indicates that neither all the high-energy flares have an
optical counterpart, like in the flaring episode occurring in 2009 January, nor all optical flares correspond to a
$\gamma$-ray flare, as in the case of the 2011 January episode.
Furthermore, the characteristics of the shocks may change
among the various flaring episodes, producing different polarization
properties as well as different light curve behaviours at the various
frequencies. \\

\section*{Acknowledgments}

Part of this work was done with the contribution of the Italian
Ministry of Foreign Affairs and Research for the collaboration project
between Italy and Japan. The VERA is operated by the National
Astronomical Observatory of Japan. This work was partially supported
by Grant-in-Aid for Scientific Researchers (24540240, MK) from Japan
Society for the Promotion of Science (JSPS).
The {\it Fermi} LAT Collaboration acknowledges generous ongoing support
from a number of agencies and institutes that have supported both the
development and the operation of the LAT as well as scientific data analysis.
These include the National Aeronautics and Space Administration and the
Department of Energy in the United States, the Commissariat \`a
l'Energie Atomique 
and the Centre National de la Recherche Scientifique / Institut
National de Physique 
Nucl\'eaire et de Physique des Particules in France, the Agenzia
Spaziale Italiana 
and the Istituto Nazionale di Fisica Nucleare in Italy, the Ministry
of Education, 
Culture, Sports, Science and Technology (MEXT), High Energy Accelerator Research
Organization (KEK) and Japan Aerospace Exploration Agency (JAXA) in Japan, and
the K.~A.~Wallenberg Foundation, the Swedish Research Council and the
Swedish National Space Board in Sweden. Additional support for science analysis during the operations phase is gratefully
acknowledged from the Istituto Nazionale di Astrofisica in Italy and the Centre National d'\'Etudes Spatiales in France.\\
This research is partly based on observations with the 100-m telescope of
the MPIfR (Max-Planck-Institut f\"ur Radioastronomie) at Effelsberg and with the
IRAM 30-m
telescope. IRAM is supported by INSU/CNRS (France), MPG (Germany) and IGN
(Spain).
This research has made use of the
data from the MOJAVE database that is maintained by the MOJAVE team
(Lister et al. 2009, AJ, 137, 3718). The OVRO 40-m monitoring program
is supported in part by NASA grants NNX08AW31G and NNX11A043G, and NSF
grants AST-0808050 and AST-1109911. Part of the research is based on
observations with the Medicina telescope operated by INAF - Istituto
di Radioastronomia. We acknowledge the Enhancement Single-Dish Control
System (ESCS) Development Team at the Medicina telescope.  \\
Data from the Steward Observatory spectropolarimetric project were
used. This program is supported by Fermi Guest Investigator grants
NNX08AW56G and NNX09AU10G.
This research has made use of the NASA/IPAC
Extragalactic Database NED which is operated by the JPL, Californian
Institute of Technology, under contract with the National Aeronautics
and Space Administration.

\end{document}